\documentclass[conference]{IEEEtran}
\usepackage{cite}
\usepackage{amsmath,amssymb,amsfonts}
\usepackage{algorithmic,algorithm}
\usepackage{graphicx}
\usepackage{subcaption}
\usepackage{textcomp}
\usepackage{xcolor}

\newcommand{\etal}{{\it et al.}}

\def\BibTeX{{\rm B\kern-.05em{\sc i\kern-.025em b}\kern-.08em
    T\kern-.1667em\lower.7ex\hbox{E}\kern-.125emX}}
\begin{document}
\title{OneStopTuner: An End to End Architecture for JVM Tuning of Spark Applications\\
}

\author{\IEEEauthorblockN{Venktesh V, Pooja B Bindal, Devesh Singhal, A V Subramanyam, Vivek Kumar}
\IEEEauthorblockA{\textit{IIIT-Delhi} \\
New Delhi, India}

}


\maketitle
\begin{abstract}
Java is the backbone of widely used big data frameworks, such as Apache Spark, due to its productivity, portability from JVM-based execution, and support for a rich set of libraries. However, the performance of these applications can widely vary depending on the runtime flags chosen out of all existing JVM flags. Manually tuning these flags is both cumbersome and error-prone. Automated tuning approaches can ease the task, but current solutions either require considerable processing time or target a subset of flags to avoid time and space requirements.

In this paper, we present \textbf{OneStopTuner}, a Machine Learning based novel framework for autotuning JVM flags. \textbf{OneStopTuner} controls the amount of data generation by leveraging batch mode active learning to characterize the user application. Based on the user-selected optimization metric, \textbf{OneStopTuner} then discards the irrelevant JVM flags by applying feature selection algorithms on the generated data. Finally, it employs sample efficient methods such as Bayesian optimization and regression guided Bayesian optimization on the shortlisted JVM flags to find the optimal values for the chosen set of flags. We evaluated \textbf{OneStopTuner} on widely used Spark benchmarks and compare its performance with the traditional simulated annealing based autotuning approach. We demonstrate that for optimizing execution time, the flags chosen by \textbf{OneStopTuner} provides a speedup of up to \textbf{1.35$\times$} over default Spark execution, as compared to \textbf{1.15$\times$} speedup by using the flag configurations proposed by simulated annealing. \textbf{OneStopTuner} was able to reduce the number of executions for data-generation by 70\% and was able to suggest the optimal flag configuration 2.4$\times$ faster than the standard simulated annealing based approach, excluding the time for data-generation.

\end{abstract}

\begin{IEEEkeywords}
JVM, Bayesian Optimization, Active Learning, Spark
\end{IEEEkeywords}

\section{Introduction}
 
 The recent rise in the use of data-centric algorithms coupled with the availability of large amounts of data has catalyzed the popularity of modern big data frameworks such as Apache Spark \cite{zaharia2012resilient}. Spark is widely adopted in the industry due to its fault tolerance and ability to perform calculations and analysis at scale. It compiles the user program into Java Bytecode and executes it over a Java Virtual Machine (JVM) \cite{jvm}. This design choice is to benefit from the years of research spent on improving the Java language and its execution. Java provides high productivity due to automatic memory management and its support for a rich set of libraries. Moreover, due to the platform-independent implementation of Java, Spark can essentially run the same bytecode on any hardware that supports a JVM.
 
 Modern JVMs are not tailored to provide optimal performance by default for all kinds of applications. Instead, they provide a plethora of configuration flags for tuning the application performance. As an example, OpenJDK version 8u144 \cite{openjdk} supports close to 700 configuration flags. Moreover, these flags are not entirely independent and often require pairing with other flags supported by the JVM. Past work has shown that JVM configuration flags can have a significant impact on the performance of big data applications~\cite{b28}. Manually tuning these flags is a daunting task and is also error-prone. Automatic tuning approaches can ease the flag selection, but at the same time, it also requires executing the application with all possible combinations of flags, thereby leading to a massive amount of data generation and long processing time for this data \cite{b13}.
 
 In this paper, we propose OneStopTuner, a simple to use and an extremely modular framework for automatically tuning JVM configuration flags. Although we use OneStopTuner to tune JVM configuration flags for Spark applications, it can easily be modified to autotune the flags supported by any other runtime solutions. Formally, OneStopTuner seeks to find the optimal configuration \(x_{opt}\) where:
\begin{equation}
  x_{opt} = arg \; {F}(\mathcal{L}(x)) \;  x \in U
 \end{equation}


 Here, $U$ is the exhaustive set of all possible configurations, $\mathcal{L}(x)$ is the objective function that represents the metric to be optimized, $F$ may be a maximizing function for metrics like throughput, or a minimizing function for metrics like runtime and latency.

\begin{figure*}[htbp]
\centerline{\includegraphics[width=148mm,scale=0.3]{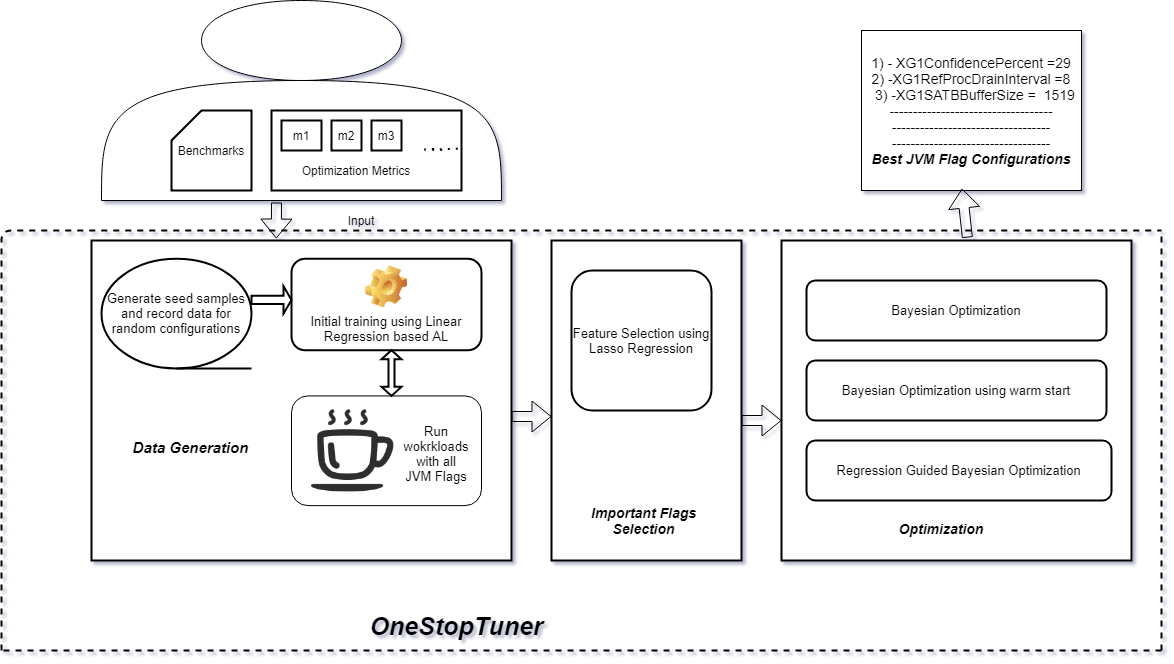}}
\caption{Overview of OneStopTuner}
\label{figBlock}
\end{figure*}
 OneStopTuner has been designed to apply its machine learning algorithms using a minimum amount of data. This is a unique feature of OneStopTuner that distinguishes it from other existing implementations. It is an essential criteria given the cost of executing jobs in a data center. OneStopTuner works in three steps.
 The first step is the data generation where the given benchmark is run iteratively for some number of times by using all the configuration flags in the JVM, but with different possible values in each iteration. OneStopTuner uses batch mode active learning in this step to capture the benchmark behaviour under different flag configurations. Data generation step is stopped once our active learning algorithm has chosen the most useful data points.
 Once the data has been generated, the second step is to apply feature selection algorithms on the generated data to shortlist the set of flags that directly impacts the user-specified optimization criteria. Third and the final step is to assign values to the shortlisted flags from step two that would help the benchmark perform best for the given optimization criteria. Here, OneStopTuner uses novel variants of Bayesian Optimization algorithms that again executes the benchmark as in the first step but only using the flags shortlisted in the second step. Bayesian optimization is well suited for finding the optimal flag configuration as we do not have an analytical function that can be optimized with respect to metrics of interest. We used two popular Spark benchmarks, Latent Dirichlet Allocation (LDA) and Dense K-Means (DenseKMeans) to evaluate the flag configuration calculated by OneStopTuner over a cluster of three nodes of 20-core Intel Xeon E5-2650 processor. We also compare the tuning performance of OneStopTuner with the widely-used simulated annealing \cite{b29} based tuning approach. We show that by using the JVM configuration flags suggested by OneStopTuner, LDA and DenseKMeans were able to achieve a speedup of upto 1.28$\times$ and 1.35$\times$, respectively, over the default Spark execution. We also show that OneStopTuner can suggest better flag combinations than simulated annealing for both these benchmarks and in significantly lesser time.

 In summary, this paper makes the following contributions:
 \begin{itemize}
 \item OneStopTuner, a Machine Learning based easy-to-use framework for automatically tuning all the configuration flags of a JVM for any given optimization criteria.
 \item A novel design for OneStopTuner that firstly uses a batch mode active learning method for application characterization, secondly it applies lasso feature selection to discard the irrelevant JVM flags. Finally, it provides three variants of Bayesian Optimization to predict the flags and the corresponding values that best suit the given optimization criteria.
 \item Evaluation of OneStopTuner by using two popular Spark benchmarks and comparing its performance with the traditional simulated annealing based autotuning approach.      
\end{itemize}        

The rest of the paper is structured as follows:
{Section II} discusses the related work. {Section III} describes the design and implementation of OneStopTuner. {Section IV} details the experimental evaluation of OneStopTuner. Finally, {Section V} concludes the paper.
    
\section{Related Work}

Tuning the parameters of an application or framework is critical for high performance and availability. However, manually tuning the flags is time-consuming and is not feasible for large applications as it is difficult to model how each flag interacts with the others. Hence, it mandates the requirement of an expert who has an in-depth knowledge of the tuning parameters and how does it affect the application behaviour.

Several attempts have been made in past across different domains to design automated algorithms to find optimal flag configurations for different software frameworks.
\cite{b16} is a comprehensive survey that explores all such attempts for tuning compiler optimization flags.
Aken \etal \cite{b14} optimized database management systems by applying Lasso analysis method with polynomial features in their regression models to find the set of important flags. On these flags they finally apply Gaussian process regression to get the optimal flag configurations.
Cereda \etal \cite{cereda2020collaborative} used a collaborative filtering based recommender system approach for exploiting similarities with previously compiled data and suggest optimized configurations for compiler systems.
Ashouri \etal \cite{b16} optimized the parameters of CPLEX (a mixed integer programming solver) by using an adaptive capping of the algorithm runs. They imposed bounds on the metrics to be optimized and terminated runs which exceed those bounds.

For big data frameworks, Sahin \etal \cite{b28} demonstrated the impact of the JVM flags on performance. However, this study was limited to flags related to heap structure and garbage collection. This is because tuning the JVM flags is not a simple task, since JVM exposes a large number of tunable flags that translates to very extensive search space. Hence, most of the studies like \cite{lengauer2014taming}, \cite{b8}, \cite{chen2002tuning}, \cite{b10} have focused on tuning a certain subset of flags and analyzing their effect on overall performance. In particular, Philipp and Hanspeter \cite{lengauer2014taming}, proposed the use of an iterated local search algorithm (ParamILS) \cite{b7} to find the best configurations for flags affecting the Garbage collection. OneStopTuner significantly differs from all these prior work as it takes into account all the configuration parameters that JVM exposes. To reduce the search space it uses a feature selection component to discard the insignificant flags.
Other works like \cite{b12} explore Guided Bayesian Optimization using a white box model that has an analytic form to find the optimal configurations for data analytics platforms. They do not face the issue of large search space as they propose to tune only memory specific JVM and Spark flags. A limitation of their approach is that a white box model necessitates expert intervention and makes the tool system dependant and non generic. Jai \etal \cite{b27} optimized Spark performance by tuning the Memory Manager configurations and were able to obtain performance improvements up to 25\%. There have been attempts even to tune JIT flags manually \cite{b4}, \cite{b5}.

Opentuner \cite{ansel2014opentuner} is another popular framework that was designed for program autotuning. It supports domain specific tuning and an ensemble of methods ranging from random search to simulated annealing methods. The random search and hill climbing methods select the values randomly, independent of the results in the previous trial runs. However, Opentuner differs from our OneStopTune as it does not support sample efficient methods like Bayesian Optimization.


For reducing the search space, JATT \cite{b13} included the entire set of available flags in their search space, but solve the infeasible complexity by defining JVM flag hierarchies by mapping the inter relations of the flags. However, their tuning process consumes a considerable duration for providing the optimal flag configurations. They leveraged the aforementioned OpenTuner \cite{ansel2014opentuner} framework for tuning the JVM flags. They reported the tuning time for SPEC benchmarks of up to 3.5 hours. Moreover, JATT does not characterize the applications and hence the search space includes flags that might have no effect on the metric of interest. Our pipeline helps address this issue by providing an option to characterize the applications and helps reduce the search space thereby reducing the time required to obtain the optimal flag configurations.

\section{Design And Implementation}

In this section, we discuss the design and implementation of OneStopTuner.

\subsection{High-level Design}
Our tool is primarily designed to be used by any developer or user for finding optimal JVM flag configurations for a given application. Figure \ref{figBlock} shows an overview of the architecture.
The first phase is data generation, and that leverages AL for application characterization.
We use Batch-mode Expected Model Change Maximization (BEMCM) Active Learning to identify the flag configurations which will provide us with the most useful data.
It is implemented as a background python script that can be triggered by the application name as input. The script runs the application repeatedly and records the metrics of interest. The AL loop stops when there is no significant improvement in validation RMSE between runs and the collected data is stored in a csv file. This portion of our pipeline is configurable and the user can characterize any application by providing an executable file or command to execute the application. Also, the user can easily extend the tool to record any desired metrics like heap usage, execution time, class load rate etc.  Our implementation of Active learning is inspired from an open source code \cite{ActiveLearningCode}.

Next, the data (flag configurations with recorded metrics) is passed through the feature selection component where the most relevant flags with respect to the desired metric are selected and saved. This is done using a background python script that runs after the application characterization phase. We use {lasso} regression \cite{b21,b22}, but our tool can be extended to use other feature selection methods as per the user's needs.

Finally, sample efficient algorithms like BO, RBO and BO with warm start are employed to find the optimum values of the configuration parameters, such that the desired metric can be optimized.
We provide a UI for ease of use. A screenshot of OneStopTuner UI is shown in Figure 2. The UI was developed using ReactJS. The backend houses the optimization algorithms mentioned earlier and are exposed through a REST API.
The user can choose from one of the several algorithms that we provide, or can extend the tool to add custom algorithms as per requirement.
We plan to release our tool as an open source software so that it can be freely used for personal or research purposes.
\begin{figure}[htbp]
\centerline{\includegraphics[height=95mm,width=80mm,scale=0.3]{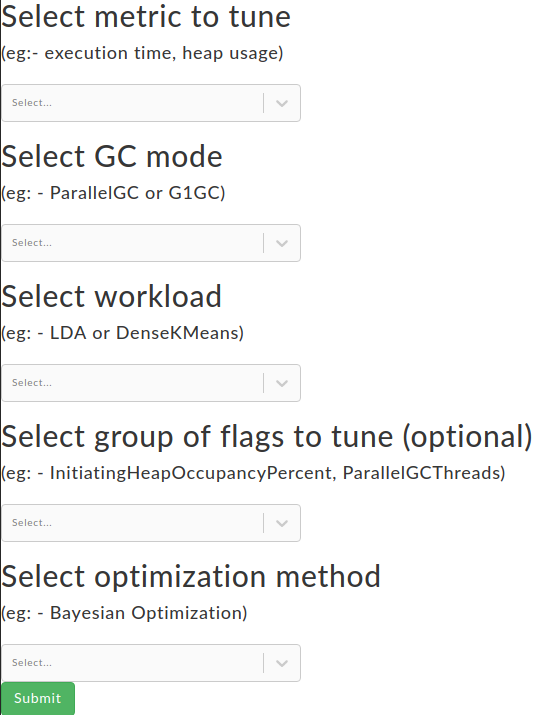}}
\caption{A screenshot of OneStopTuner UI}
\label{fig}
\end{figure}

In the following sections, we provide an overview of each component of our tool.

\subsection{Application Characterization using Active Learning}
The first step of our pipeline is to characterize the given application. By characterization we mean capturing the runtime behaviour of a given application under different configurations on a particular machine setup. Most flag configurations have a direct impact on the metrics recorded. For instance, flags like \textit{InitiatingHeapOccupancyPercent} has an impact on {execution time} as tuning it correctly can help avoid full GC cycles. This reduces the step to a simple task of setting the configuration flags to certain values (random values for data generation), and recording the resulting behaviour of the application execution. However, the total number of flags is about 700 for JVM version 1.8.0\_144 \cite{jvm}, and most of these flags may take a continuous set of values. Even if the flags had all been binary, we would have required more than $2^{700}$ flag configurations to completely map the search space. To be able to accurately train on this massive search space, we would require a large number of data points.
For heavy applications, this is a huge overhead as several thousand runs may take days. OneStopTuner minimizes the data generation time by reducing the amount of data that is required.
We do this by using Active Learning (AL), a technique for learning under limited labels. Using AL, we can find the configuration values, such that if we obtain the metrics for those values of configurations we will tune the weights of our model the most. Hence, our model can converge on much lower amounts of data than would be normally required.

One of the popular methods in AL is uncertainty sampling \cite{b18,b19}. This method chooses the samples about which the model is the most uncertain and queries the ground truth label. In our case, instead of querying a table with the correct data pre recorded, we run the application and collect the real desired data.
Query By Committee (QBC) \cite{b20} is another popular method where an ensemble of models is trained and the samples on which the models disagree the most are chosen.
However most of such active learning methods are sequential in nature. In every sampling iteration one informative sample is chosen and model is retrained for each informative sample which is computationally expensive. To leverage parallelism and retrain the data on a batch of informative samples, we have the Batch Mode Active Learning (BMAL) algorithm. BMAL selects a batch of samples in each iteration which helps in gathering more data in lesser number of iterations. BMAL, however, does not consider the similarity or correlation between selected samples. This could potentially be a significant problem for us since if we select several correlated data points, we get a lot of redundant flag configurations that are not informative. To overcome this issue, we adopt the approach proposed in \cite{b3} Batch Mode Expected Model Change Maximization (BEMCM). BEMCM is a unified framework for applying active learning for regression. BEMCM chooses examples that change the model's parameters the most. It is based on the intuition that samples that update the model parameters by a significant amount help in generalization and are more informative. The change is calculated as the norm of the gradient at a single candidate example. BEMCM extends EMCM to select $k$ samples at each iteration that best matches the performance of sequential EMCM when performed for $k$ iterations.

 \begin{algorithm}
 \caption{BEMCM for application characterization}
 \label{algoBEMCM}
 \begin{algorithmic}
 \renewcommand{\algorithmicrequire}{\textbf{Input:}}
 \renewcommand{\algorithmicensure}{\textbf{Output:}}
 \REQUIRE A small initial set $D=\{j_i,y_i\}_{i=1}^n$ where $j$ - JVM flag configurations, $y$ - metrics like execution time or heap usage, the unlabelled set $U=\{j_{i}^*\}_{i=1}^m$, size of ensemble $Z$, batch size $k$, linear regression model $f(j)$ initially trained on $D$
 \ENSURE  batch of informative samples, $i=\{j_{1}^*,j_{2}^*...j_{k}^*\}$
  \STATE initialize $i=\phi$
  \STATE $B(Z)=\{f_1,f_2..,f_Z\}$  \COMMENT{ensemble of models generated by bootstrap}
  \WHILE{$|i|<k$}
  \FOR{each $j^*$ in $U$}
   \STATE $\{y_1...y_Z\} \gets B(Z)$
   \STATE Estimate the change of model parameters
  \ENDFOR
  \STATE Select $j^*$ leading to the maximum change of model parameters 
  \STATE $i \gets i \bigcup j^*$
  \ENDWHILE
 \RETURN $i=\{j_{1}^*,j_{2}^*...j_{k}^*\}$
 \end{algorithmic} 
 \end{algorithm}
 
Algorithm \ref{algoBEMCM} details the steps of application characterization using active learning. In our implementation of BEMCM, the metrics are considered as target variables and the flag configurations are considered as features. We train a {linear regression} model with polynomial features on a seed set of samples generated by running the application several times and then the active learning loop is initiated. In the active learning loop, at each iteration, we choose the samples that change the model parameters the most and query their labels by running the application to record the runtime statistics as described earlier. Then the new data is added to training data and the model is retrained. In order to determine the most informative samples, we have to estimate the change of model parameters to choose the samples that maximize this change. As we use stochastic gradient descent learning rule for training the model, the change can be approximated as the gradient of the loss function with respect to candidate JVM flag configurations. We adapt the derivation proposed by Cai \etal \cite{b3} to our scenario. 

Consider a linear regression model 
\begin{equation}
    f(j;W)=W^\top j
\end{equation}
The objective of training linear regression is to reduce the squared error between the predicted metric and actual metric.
\begin{equation}
      E(W)=\frac{1}{2}\sum_{i=1}^{n}(f(j_{i})-y_{i})^{2}
\end{equation}
   where $j$ denotes the JVM flag configurations and $y_i$ is the ground truth metric (for example, execution time) for the given flag configuration.
 When a new set of JVM flag configuration $j^*$ with $y^*$ as label is added as a candidate to the dataset, the error on new dataset becomes:
 \begin{equation}
           E(W)=\frac{1}{2}\sum_{i=1}^{n}(f(j_{i})-y_{i})^{2} + L_{j^*}(W)
 \end{equation}
 where,
 \[L_{j^*}(W)=\frac{1}{2}(f(j_{i}^*)-y_{i}^*)^{2}\]

The derivative for squared loss $L_{j^*}(W)$ with respect to parameters $W$ at $j^*$ is:
\begin{equation}
\begin{aligned}
    \frac{\partial L_{j^*}(W)}{\partial W}&=(f(j^*)-y^*) \frac{\partial f(j^*)}{\partial W} \\
    &=(f(j^*)-y^*) \frac{\partial W^\top j^*}{\partial W} \\
    &=(f(j^*)-y^*)j^*
    \end{aligned}
\end{equation}

The true label $y^*$ is not known in advance and hence we employ bootstrap to create an ensemble $B(Z)=\{f_1,f_2..,f_Z\}$ that estimates the prediction distribution. The estimated model change approximates the true model change. Then the informative sample is chosen as the one that leads to the highest model change.

This method obviates the need for a large number of runs to characterize the application and also the algorithm requires fewer iterations than the sequential active learning counterparts.

\subsection{Selecting the most important flags}
The next step in the pipeline is to select the most important flags. We want to do this to reduce the feature set for our tuning algorithms, so that they can converge sooner and on lesser data.
Here, we use the Lasso Regression algorithm, but the user can easily extend the tool to use any custom algorithm. The user may also decide to skip the feature selection entirely, if they feel they can train comfortably on the entire flag set or if they have manually selected the best flags. If feature selection is not to be performed, then the AL loop of data generation may also be skipped, and the user may directly apply the tuning algorithm. Do note that in that case, BO with warm start will be unavailable, since that relies on reusing data collected during the first phase.
We choose to apply lasso regression due to its ability to induce sparsity in the solution. This will lead to selection of only important flags.
Lasso regression tries to minimize the following objective,
\begin{equation}
\min_w \;\lVert y - jw \rVert^2_2 + \lambda \lVert w\rVert^2_1
\end{equation}
            
where $w$ denotes weights, $j$ denotes JVM flags, $y$ denotes the metric to be optimized (execution time or heap usage) and $\lVert . \rVert$ denotes norm.
Through this, we are able to retain only the flags that have a strong correlation to the metric under discussion. 


\subsection{Recommending the best flag configurations}
After selecting the most important flags, our tool identifies the best JVM flag configuration values to optimize the system performance. For this, we investigate Bayesian Optimization (BO) and implement the widely used Simulated Annealing (SA) method as a baseline. We also propose two variants of BO that leads to recommendation of better JVM flag configurations and faster convergence.
We use Bayesian Optimization \cite{b23} since it is a sample efficient technique, meaning that it can be used to train on data where the number of samples are limited and difficult to collect. In our scenario, trying out various flag configurations for a large number of times is infeasible and expensive from a time consumed view. Thus, Bayesian Optimization is extremely useful for our purpose to reduce the data required.

Bayesian Optimization solves optimization problems where the objective function does not have an analytic expression but rather can be evaluated only through time expensive experiments. It incorporates an assumed or semi known prior belief about the objective function and updates the prior by sampling the real values of the target metric. After many such iterations, it is able to form an accurate posterior function which can represent the objective function that had to be learned. The model used for approximating the objective function initially is called surrogate model. BO also uses an acquisition function that determines the candidates to sample that are likely to improve the objective function the most.

Gaussian process (GP) is a good surrogate as it provides good uncertainty estimates (confidence intervals) and is analytically tractable. We propose to use {Expected Improvement} as an acquisition function. Expected improvement is defined as:
\begin{equation}
I(x) = \max(0, f(x) - f(x^*))
\end{equation}

The expected improvement (EI) acquisition function chooses the point $x$ that maximizes the expected value of $I(x)$ under the GP posterior. The goal is to determine whether sampling $x$ would produce a new value $f(x)$ that is better than the current best $f(x^*)$ value. 
 \begin{algorithm}
 \caption{Bayesian Optimization}
 \label{algoBO}
 \begin{algorithmic}
 \renewcommand{\algorithmicrequire}{\textbf{Input:}}
 \renewcommand{\algorithmicensure}{\textbf{Output:}}
 \REQUIRE Initial configuration generated using SOBOL $D$, \\
       \hspace{18pt}   Configurations range $R$, \\
       \hspace{18pt}   Objective function $Q$
 \ENSURE  Optimal configurations
  \STATE fit surrogate model on initial data
 \WHILE {not convergence}
    \STATE Let $x_m$ be the configuration that maximizes the acquisition function
    \STATE  $y_m \gets Q(x_m)$
    \STATE add $x_m$ and $y_m$ to the dataset and fit surrogate model on new data
\ENDWHILE
 \RETURN flag configurations with largest or smallest $Q(x)$ depending on metric of interest 
 \end{algorithmic} 
 \end{algorithm}
 
Algorithm \ref{algoBO} details the steps of BO. We generate initial samples using quasi-random SOBOL sequence \cite{b24} for exploration. After a few iterations, the Gaussian process is fit to the generated points and BO loop starts. Here, the data of flag configurations and metrics collected in the data generation phase using AL are not used.
 

We now describe our the two BO variants for obtaining the best JVM flag configurations:
\begin{itemize}
    \item \textbf{BO with warm start}: Leveraging data from previous runs would help the BO method to explore more useful parts of the search space. Hence we investigate a variant of BO, where the data generated using AL from the application characterization phase is used. This variant of BO is derived from Algorithm \ref{algoBO} by replacing the quasi-random samples (SOBOL) with data collected using AL in the Input section. The BO optimization is informed by the Gaussian process model trained on the data. We observed that this variant leads to better configurations for most applications.
    
    \item \textbf{Regression Guided BO (RBO)}: A major component that consumes time in the above BO methods is the evaluation function $Q$ that runs the Spark application. Hence the total tuning time equals the number of BO iterations times the application execution time. For heavy applications, it is desirable to get a flag configuration that is better than default execution time without incurring the cost of having to run the application. Hence we propose a novel variant of BO where instead of running the application to evaluate the chosen flag configurations, we use a prediction model to predict the metric. This variant can be derived from Algorithm \ref{algoBO} by replacing the function $Q$ with the predictor. We use the linear regression model trained using \textit{AL} during data generation phase. We observe that the predictions produced were close to actual performance in most cases. RBO provides a speedup of factor 6$\times$ compared to BO methods with a slight trade-off in accuracy and is suitable for heavy applications.

\end{itemize}



\begin{table}[htbp]
\caption{Benchmark applications used in evaluation}
\begin{center}
\begin{tabular}{|c|c|}
\hline
\textbf{Application} & \textbf{Dataset} \\
\hline
Latent Dirichlet Allocation  & Hibench LDAExample, large,\\&
 10000 documents,maxResultSize 3GB  \\
\hline
Dense K-Means & DenseKMeans, Hibench, large, \\& 
20M samples, 20 dimensions \\
\hline

\hline
\end{tabular}
\label{tab1}
\end{center}
\end{table}

 

\begin{figure*}
\centering
\captionsetup{justification=centering}
\begin{subfigure}{0.21\linewidth}
\centering\includegraphics[width=1\linewidth]{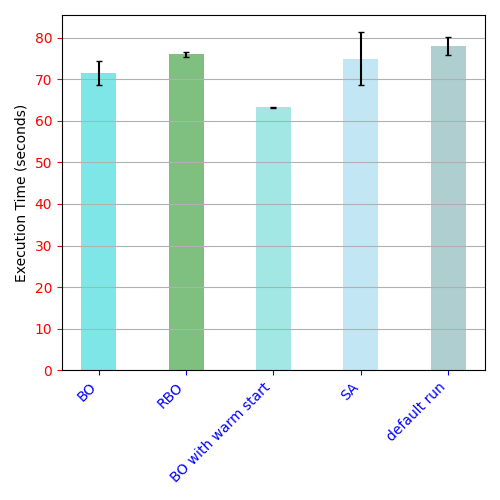}
\caption{LDA (ParallelGC)}
\end{subfigure}%
\hspace{2em}
\begin{subfigure}{0.21\linewidth}
\centering\includegraphics[width=1\linewidth]{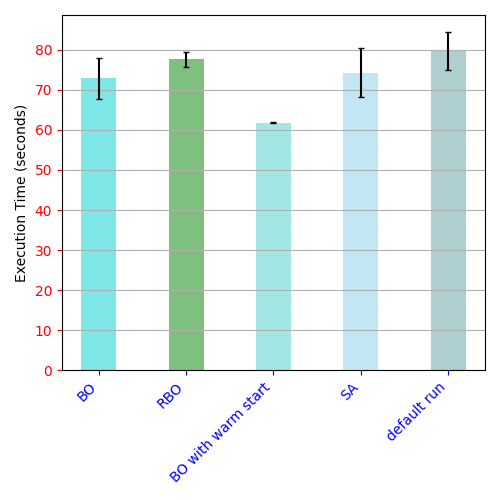}
\caption{LDA (G1GC)}
\end{subfigure}%
\hspace{2em}
\begin{subfigure}{0.21\linewidth}
\vspace{8pt}
\centering\includegraphics[width=1\linewidth]{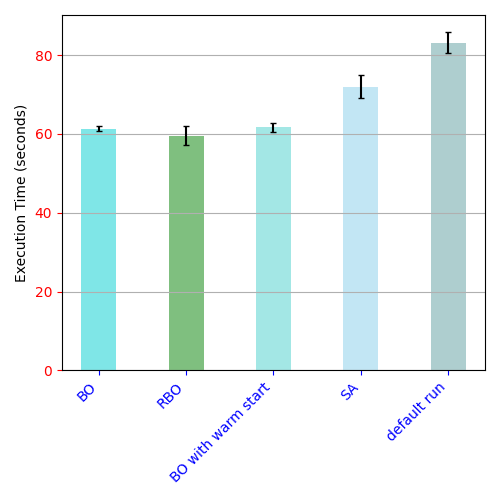}
\caption{{DenseKMeans (ParallelGC)}}
\end{subfigure}
\hspace{2em}
\begin{subfigure}{0.21\linewidth}
\centering\includegraphics[width=1\linewidth]{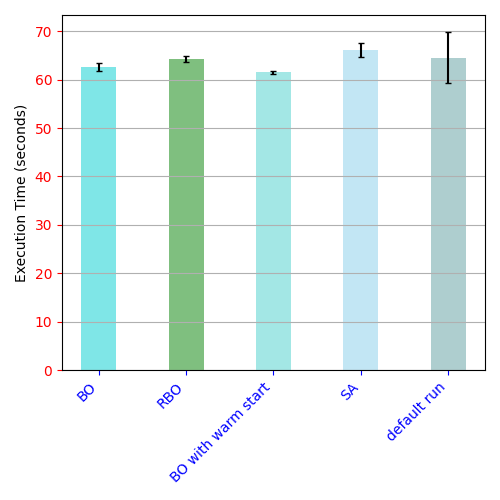}
\caption{{DenseKMeans (G1GC)}}
\end{subfigure}%
\caption{Execution Times for benchmark runs using default configuration values, and using configuration values tuned by OneStopTuner}
\label{fig}
\end{figure*}

\section{Experimental Methodology}
In this section we describe our experimental setup. We used two benchmarks from HiBench suite \cite{b1} for evaluating OneStopTuner. The details of these benchmarks are listed in Table I. We carried out all our experiments on a local cluster of three nodes of dual socket Intel Xeon E5-2650 processor running at 2.3GHz. Each socket has a total of 10 physical cores, i.e., total number of cores in our cluster is 60. Memory available per node is 90GB. We used Java HotSpot (TM) 64-Bit Server Virtual Machine version number 1.8.0\_144.

The following subsections detail the experiments performed.
\subsection{Data Generation}
For characterizing the benchmarks we first generated an initial dataset using random flag configuration samples and recording the target metric (labels) by running the benchmark for 30\% of the samples. Then 20\% of the labelled samples were allotted as test set and 10\% as initial seed set for the linear regression model. Rest 70\% of the data was designated as an unlabelled set. The AL loop was executed for 10 iterations. In each round of sampling, about 3\% of unlabelled set are selected and labelled. The active learning loop is stopped when there is no change in the Root Mean Square Error (RMSE) between Active learning rounds. Another candidate for the labeling budget could be the maximum time allowed for the data generation phase. Data was generated for each benchmark individually by using 3 Spark executors (one executor at each node).

\subsection{Metrics Recorded}
\label{evaluationMetrics}
For our experiments, we chose \textbf{execution time} and \textbf{heap usage} as optimization metrics, and hence recorded these metrics while generating the data.
Heap usage percentage ($HU$) is recorded using a method  similar to the one proposed in JATT \cite{b13}. HU is calculated from jstat statistics as:
\begin{equation}
HU = \dfrac{{S0U+S1U+EU+OU}}{ S0C+S1C+EC+OC} \times 100
\end{equation}
 where, $S0U$: Survivor space 0 utilization, $S1U$: Survivor space 1 utilization, $EC$: Current eden space capacity, $EU$: Eden space utilization, $OC$: Current old space capacity, $OU$: Old space utilization.
The above statistics are recorded for every 5 seconds and hence the statistics are recorded several times for a single run of benchmark. The average heap usage for a single run is calculated as,
\begin{equation}
\dfrac{\sum_{k=1}^n HU_k}{n}
\end{equation}
The execution time is recorded directly by timing the run.
\subsection{Feature Selection}
For selection of the most important flags we use {Lasso Regression} \cite{lasso} from sklearn . We set the value of the hyperparameter, $\lambda$ = 0.01 using grid search. 

\subsection{Tuning Experiments}
The benchmarks mentioned in Table I were executed for two GC modes G1GC and ParallelGC. However, our tool supports tuning with all GC modes. We group the flags according to GC modes as flags that are specific to a particular GC mode cannot be tuned when using another GC mode. For instance,  the flag \textit{G1HeapRegionSize} is relevant when using G1GC but not ParallelGC. We extract the list of JVM flags using the command \textit{java -XX:+PrintFlagsFinal} and group the flags according to GC modes similar to JATT \cite{b13}. In addition to the GC flags, we also tune compiler related flags and other common flags. This is to offer flexibility in tuning and as observed in \cite{b13}, tuning all flags without restricting to a subset of flags helps in improving the performance further.

We ran tuning experiments for each of the benchmark individually using the entire cluster resources. Since in industrial deployments applications run in parallel, we also ran tuning experiments by running both the benchmarks in parallel and report the results in Section \ref{results}.
All tuning experiments were repeated 10 times to avoid random fluctuations and ensure that the performance gains are consistent. The mean and standard deviation of the performance metric across 10 runs are also plotted.

For implementing BO and its variants (BO with warm start and RBO), we use the AX-platform \cite{b26}.
We observed that BO and variants mostly converge in 20 iterations and hence a single tuning run involves 20 iterations for all the benchmarks.

\subsection{Simulated Annealing} We consider Simulated Annealing (SA) as one of the baseline methods as it is a popular approach used in several prior works \cite{ansel2014opentuner}, \cite{b13}. We were unable to directly use any open-source implementations (e.g., Opentuner) for comparing OneStopTuner as we found it wasn't supported on latest Java and Spark versions. We used Latin Hypercube sampling (LHS) \cite{b25} of SA. Latin Hypercube Sampling is a general way of generating random samples of parameter values, which enables the SA method to explore more useful flag configurations.
The main motivation behind applying LHS is that it is empirically proven to be useful in cutting down processing time \cite{b25} by more than 50 percent. This would enable a fair comparison between Bayesian Optimization and LHS as in both the scenarios, our goal is to find the best configuration for the flags using limited set of trials. 
We leverage mlrose \cite{mlrose} IDAES package\cite{IDAES} to implement Simulated Annealing in OneStopTuner.



\begin{figure}
\centering
\captionsetup{justification=centering}
\centering\includegraphics[width=.8\linewidth, height=6cm]{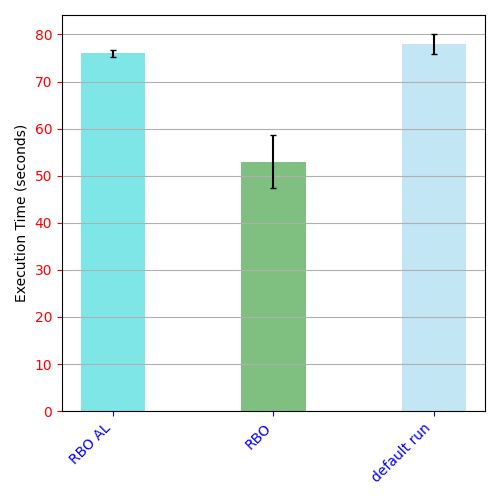}
\caption{ Comparison of RBO tuning results (execution time)}
\label{fig4}
\end{figure}

\begin{figure}
\centering
\captionsetup{justification=centering}
\centering\includegraphics[width=.8\linewidth, height=6cm]{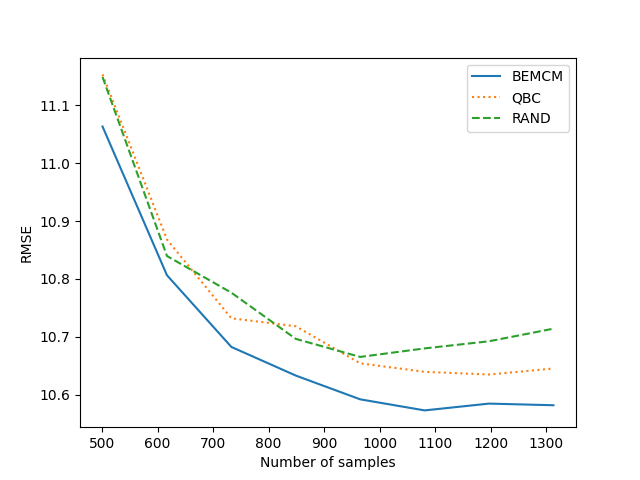}
\caption{ Plot showing the RMSE for BEMCM AL vs Linear regression (target - execution time)}
\label{fig}
\end{figure}

\begin{figure*}
\centering
\begin{subfigure}{0.24\linewidth}
\includegraphics[width=1\linewidth]{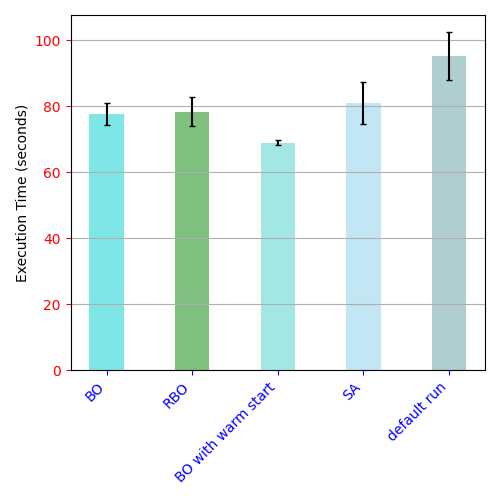}
\caption{{LDA-G1GC}}
\end{subfigure}
\begin{subfigure}{0.24\linewidth}
\includegraphics[width=1\linewidth]{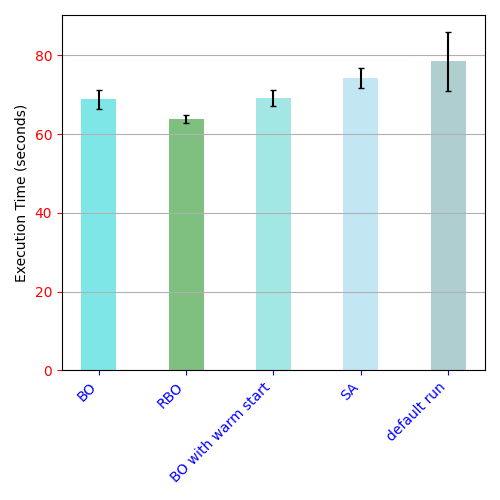}
\caption{DenseKMeans-G1GC  }
\end{subfigure}%
\label{subfig}
\begin{subfigure}{0.24\linewidth}
\includegraphics[width=1\linewidth]{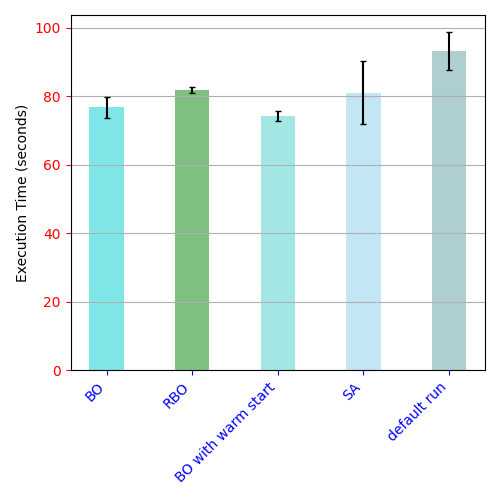}
\caption{{LDA-G1GC   }}
\end{subfigure}
\begin{subfigure}{0.24\linewidth}
\includegraphics[width=1\linewidth]{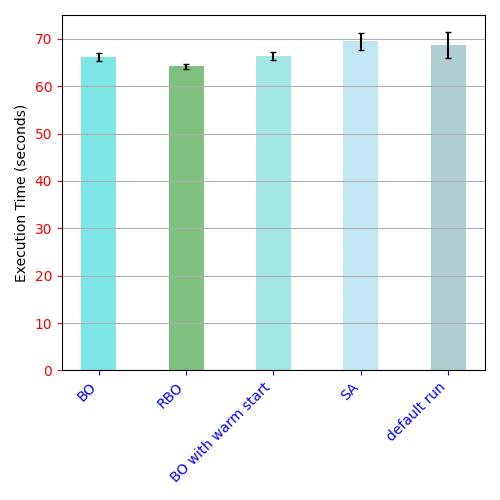}
\caption{DenseKMeans-G1GC }
\end{subfigure}%

\caption{ Tuning results for execution time with benchmarks running in parallel with 2 executors, 15 cores, 60GB per executor (a, b) and  with 3 executors, 10 cores each executor (c, d)}
\label{fig}
\end{figure*}

\begin{figure*}
\centering
\begin{subfigure}{0.24\linewidth}
\includegraphics[width=1\linewidth]{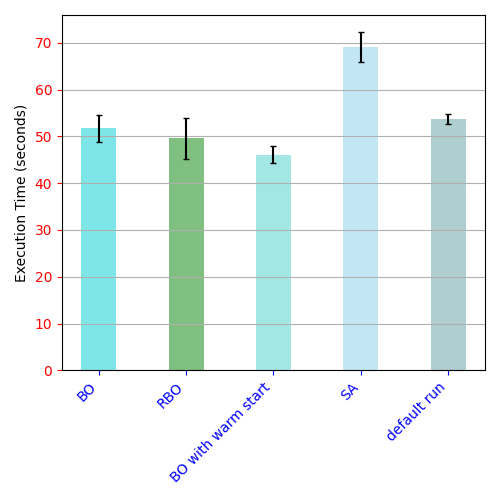}
\caption{{LDA-ParallelGC }}
\end{subfigure}%
\begin{subfigure}{0.24\linewidth}
\includegraphics[width=1\linewidth]{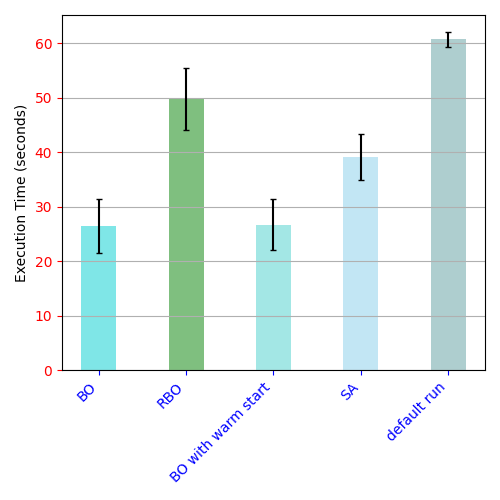}
\caption{LDA-G1GC  }
\end{subfigure}
\begin{subfigure}{0.24\linewidth}
\includegraphics[width=1\linewidth]{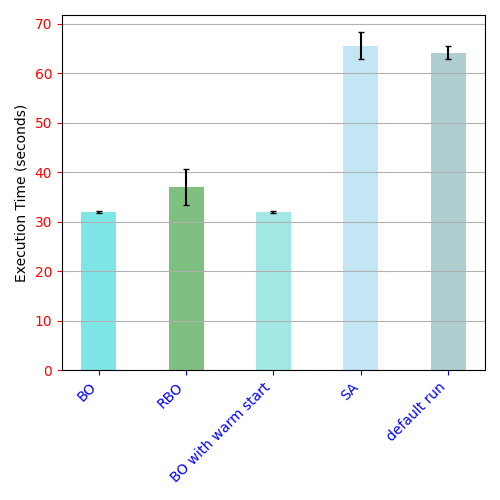}
\caption{{DenseKMeans-ParallelGC }}
\end{subfigure}
\begin{subfigure}{0.24\linewidth}
\includegraphics[width=1\linewidth]{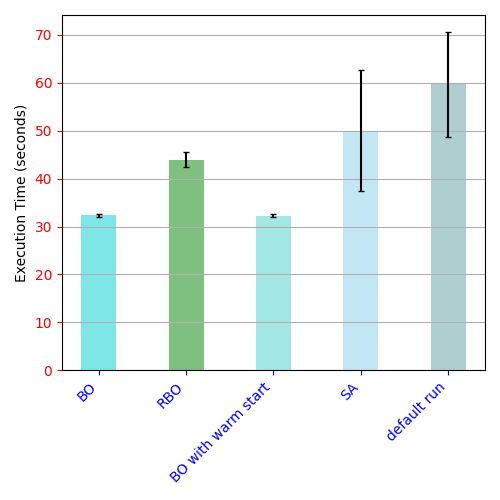}
\caption{{DenseKMeans-G1GC  }}
\end{subfigure}
 
\caption{ Heap Usage tuning results }
\label{fig}
\end{figure*}

\begin{table}[ht]
    \caption{Flags selected by lasso regression for LDA and DenseKMeans (DK)}
    \centering
    \begin{tabular}{|c|c|c|}
    \hline benchmark& { \# of flags exec. time} & {\# of flags heap usage} \\
    \hline
    LDA, ParallelGC & 99 & 101\\
    LDA, G1GC & 108 & 117\\
    DK, ParallelGC& 100& 96\\
    DK, G1GC& 97&107 \\\hline
    \end{tabular}
    \label{tab:flagsselected}
\end{table}
\section{Experimental Evaluation}
\label{results}
In this section we present the result of experimental evaluation of OneStopTuner. 

\subsection{Benefit of using Feature Selection} 
We first describe the benefit of using feature selection on flag pruning. The results of the flags selected for different metrics using feature selection are shown in Table \ref{tab:flagsselected}. The default number of flags grouped under ParallelGC mode and G1GC mode are originally 126 and 141 respectively. The table shows that the feature selection helped OneStopTuner in significantly reducing the number of flags to be tuned when optimizing for a given metric.


\subsection{Benefit of using Active Learning} In this section, we discuss the significance of active learning and demonstrate the results obtained when using BEMCM AL \cite{b3} method (Figure \ref{figBlock}).

We compared tuning results of RBO for two Linear Regression models Figure \ref{fig4}. The comparison is reported only for the LDA, but similar trend was observed even for DenseKMeans. One model was trained on 2000 samples collected without leveraging AL and the other LR model was trained on 600 samples generated through AL. Data generation through AL required lesser number of samples as only the most informative samples are chosen at each AL round. It is evident that the model trained through BEMCM AL method produces better results as the predicted values are closer to the actual execution time.

The plot in Figure 5 shows the change in RMSE as more samples are acquired. We demonstrate that the Linear Regression (LR) model trained using BEMCM method converges faster than LR trained with random selection of data and LR model trained using QBC AL method. This confirms our intuition that model trained using BEMCM AL method has better generalization capabilities. This is because the samples that update the model parameters the most are considered as informative samples. RBO leverages the LR model trained to predict the metric of interest.

\subsection{Time taken to tune}
Owing to using more sample efficient methods, OneStopTuner can converge faster than other tools which implement Simulated Annealing. All the tuning done as a part of our experiments was done over 20 iterations each, and we noticed that for LDA (G1GC), OneStopTuner was able to execute the 20 iterations in 1850 seconds (or 30.83 minutes) whereas Simulated Annealing took 2914 seconds (or 48.56 minutes). This shows that our tool is 1.57$\times$ faster than Simulated Annealing algorithms.
Similarly, for DenseKMeans (G1GC), OneStopTuner took 1294 seconds (or 21.56 minutes) to tune, whereas Simulated Annealing took 3124 seconds (or 53.36 minutes), showing a speed factor of 2.41$\times$.

\subsection{Tuning Result for Execution Time}

We now describe the performance gains obtained by each benchmark using the JVM flags tuned by OneStopTuner. Table \ref{tab:ExecutionSpeedupTable} shows the execution time speedups (over default execution times) that are obtained for different benchmarks via different algorithms.

\begin{table}[ht]
    \caption{Execution Time speedups over default times, for LDA and DK(DenseKMeans)}
    \centering
    \begin{tabular}{c|c|c|c|c}
    \hline Benchmark, GC & BO & RBO & BO, warm start & SA\\
    \hline
    LDA, ParallelGC & 1.09$\times$ & 1.03$\times$ & \textbf{1.23$\times$} & 1.04$\times$\\
    LDA, G1GC & 1.09$\times$ & 1.02$\times$ & \textbf{1.28$\times$} & 1.07$\times$\\
    DK, ParallelGC & 1.36$\times$ & \textbf{1.39$\times$} & 1.35$\times$ & 1.15$\times$\\
    DK, G1GC & 1.02$\times$ & 1.0$\times$ & \textbf{1.04$\times$} & 0.97$\times$\\
    \end{tabular}
    \label{tab:ExecutionSpeedupTable}
\end{table}

\subsubsection*{LDA}
Figure 3a shows the tuning results in ParallelGC mode for LDA benchmark. We observe a speedup of 1.23$\times$ over the default execution times when using BO with warm start, where the Gaussian process surrogate is trained on the data generated during the data generation phase. We observe that in Figure 3a, BO with warm start outperforms all other methods. Though BO initiated with SOBOL sequence provides a speedup of 1.08$\times$ over the default configurations, the performance improvement provided by the warm start variant is significant. We show that the informative samples collected in data generation using AL actually provide better prior knowledge of the objective to be optimized. This helps the optimization procedure to explore more useful regions of the parameter space. 

The performance of LDA, in case of G1GC mode is shown in Figure 3b. We can observe that the BO with warm start gives a speedup of {1.27$\times$} over the default run.

\subsubsection*{DenseKMeans}
Figure 3c shows the tuning results in ParallelGC mode for DenseKMeans benchmark. We observed a speedup of 1.35$\times$ over the default execution times when using BO with warm start. This is higher than the improvement observed in LDA. That is because the input size to DenseKMeans is 72GB which is split across 1915 tasks. This causes frequent long GC pauses and hence tuning flags like \textit{InitiatingHeapOccupancyPercent} and other flags can help avoid \lq\lq{stop the world}\rq\rq\ GC cycles. 

In case of G1GC mode, we can observe that though the performance increases as evident from Figure 3d, the speedup is of order 1.04$\times$. This is because G1GC avoids long GC pauses and hence the default run here is better than the default run in ParallelGC mode.


\subsection{Parallel Run} Figure 6 presents the results of tuning experiments with both LDA and DenseKMeans benchmarks running in parallel. This better mirrors real time industrial scenarios and demonstrates the robustness of our tuning algorithms.
The tuning of DenseKMeans and LDA benchmarks were carried out in parallel (both benchmarks running together) in the cluster with 2 Spark executors for each benchmark, 15 cores for each Spark executor and 60GB for each Spark executor.
 
From Figure 6a it is evident that BO with warm start offers a speedup of 1.37$\times$ over default configuration. Even BO offers a speedup of over 1.2$\times$. The speedup here is significant when compared to the tuning results when LDA was tuned individually using the full cluster resources. This is because LDA benchmark runs faster with more parallelism and the performance degrades a little when the number of cores allotted are less. This indicates that proper tuning of JVM flags in such resource constrained scenarios can help improve the performance significantly and our tool is able to achieve a significant speedup 
 
The speedup gained for DenseKMeans G1GC mode in parallel setting follows a similar trend as individual tuning run of DenseKMeans in G1GC mode, as shown in Figure 6b.
 
Figure 6c and Figure 6d show the results of tuning in a different setting. Both DenseKMeans and LDA benchmarks were tuned in parallel with 3 executors for each benchmark, 10 cores for each executor and 44GB of memory for each executor for the LDA benchmark and 50GB of memory for each executor for DenseKMeans benchmark.  For LDA benchmark in G1GC mode, a speedup of 1.21$\times$ using BO and 1.25$\times$ using BO with warm start is obtained as shown in Figure 6c. DenseKMeans benchmark tuned in G1GC mode speedup gained using BO with warm start and BO have speedups of 1.03$\times$ and 1.04$\times$ over the default run as shown in Figure 6d.

\subsection{Tuning Result for Heap Usage} Table \ref{tab:HeapUsageTable} shows the heap usage optimizations that are obtained for different benchmarks via different algorithms.

\begin{table}[ht]
    \caption{Heap Usage Improvements over default usage}
    \centering
    \begin{tabular}{c|c|c|c|c}
    \hline benchmark, GC & BO & RBO & BO, warm start & SA\\
    \hline
    LDA, ParallelGC & 3.78\% & 7.83\% & 14.31\% & \textbf{28.55}\%\\
    LDA, G1GC & 56.41\% & 18.04\% & \textbf{55.94}\% & 35.51\%\\
    DK, ParallelGC & 50.13\% & 42.22\% & \textbf{50.25}\% & 2.22\%\\
    DK, G1GC & 45.86\% & 28.37\% & \textbf{45.89}\% & 16.19\%\\
    \end{tabular}
    \label{tab:HeapUsageTable}
\end{table}

The results of tuning for overall heap usage percentage is presented in Figure 7. Tuning for low memory footprint is common as it is desirable to reduce the cost incurred on virtual machines. However the heap occupancy cannot be lower than the minimum size needed for running the benchmark. We enforce this constraint by defining this in the range of the heap size related flags.  We report the results in terms of percentage improvement. We observe that the Heap Usage percentage of ParallelGC is lower when compared with G1GC for default run. However the tuning results give dramatic improvements as presented in the figure. BO and BO with warm start give improvements of 56\% and 55\% respectively as shown in Figure 7b. This indicates that the tuning of flags has lead to much smaller memory footprint. However, tuning for small memory footprint it may lead to worse configurations, that may end up slowing down the application.

\section{Conclusion}
In this paper, we present the design of {OneStopTuner}, an end-to-end pipeline that addresses the challenge of improving the performance of Spark applications.  The tool uses active learning for application characterization such that a favorable amount of data can be generated to compute the optimal flag configuration. Further, the lasso regression is used to identify relevant flags from the generated data. OneStopTuner then applies sample efficient tuning methods like Bayesian Optimization and its variants to determine the optimal values of these flags. OneStopTuner achieves an execution time speedup of upto {1.35$\times$} and upto {50\%} improvement in heap usage. Our tool is extremely modular and can be easily extended to include more optimization methods. We plan to further explore methods that can efficiently work with other metrics of interest while optimizing a given application.

\bibliography{myBib}
\bibliographystyle{plain}

\end{document}